# Shannon Perfect Secrecy in a Discrete Hilbert Space

Randy Kuang
Quantropi Inc.
Ottawa, Canada
randy.kuang@quantropi.com

Nicolas Bettenburg
Quantropi Inc.
Ottawa, Canada
nicolas.bettenburg.@quantropi.com

*Abstract*— **The One-time-pad (OTP) was mathematically proven to be perfectly secure by Shannon in 1949. We propose to extend the classical OTP from an n-bit finite field to the entire symmetric group over the finite field. Within this context the symmetric group can be represented by a discrete Hilbert sphere (DHS) over an n-bit computational basis. Unlike the continuous Hilbert space defined over a complex field in quantum computing, a DHS is defined over the finite field GF(2). Within this DHS, the entire symmetric group can be completely described by the complete set of n-bit binary permutation matrices. Encoding of a plaintext can be done by randomly selecting a permutation matrix from the symmetric group to multiply with the computational basis vector associated with the state corresponding to the data to be encoded. Then, the resulting vector is converted to an output state as the ciphertext. The decoding is the same procedure but with the transpose of the pre-shared permutation matrix. We demonstrate that under this extension, the 1-to-1 mapping in the classical OTP is equally likely decoupled in Discrete Hilbert Space. The uncertainty relationship between permutation matrices protects the selected pad, consisting of M permutation matrices (also called Quantum permutation pad, or QPP). QPP not only maintains the perfect secrecy feature of the classical formulation but is also reusable without invalidating the perfect secrecy property. The extended Shannon perfect secrecy is then stated such that the ciphertext C gives absolutely no information about the plaintext P and the pad.**

*Keywords—Shannon, perfect secrecy, OTP, permutation, quantum permutation gates, symmetric group, QPP, entropy, bijection, Hilbert space, Hilbert sphere, Bloch sphere*

## I. Introduction

The well-known Vernam Cipher was named after Gilbert Sandford Vernam, who invented the polyalphabetic stream cipher in 1917 and later filed a patent titled "signalling secret system" on September 13, 1918—also called one-time pad (OTP). The grant, with U.S. Patent number 1310719, was issued on July 22, 1919 [1]. Vernam's invention is quite possibly one of the most important in cryptographic history.

OTP is a cryptographic technique for encryption and decryption of data that cannot be cracked; however, it does require that the pre-shared true random key be the same size as the input data (called "plaintext"). The plaintext P is encrypted with the secret random key K by applying XOR operations to produce a ciphertext C. The ciphertext C can then be decrypted with the same secret key K by applying XOR operations to produce the plaintext P. Claude Shannon in 1949 proved mathematically, in his "Communication Theory of Secret Systems" [2], that the OTP exhibits the property of "perfect secrecy" - that is, the ciphertext C gives absolutely no information about the plaintext P. However, the true random key K (also called "pad") can only be used once in order to maintain the perfect secrecy property of the cryptographic process.

Although OTP provides mathematically proven perfect secrecy, OTP remains impractical for reasons that it requires secure true random generation and exchange. In recent years, we have developed robust, industry-grade techniques for generating true random number using quantum processes, for instance a quantum random number generator (QRNG) [3]. However, key exchange remains the subject of much active research activity. Alternative public key exchanges, such as RSA [4], Diffie-Hellman (DH) [5], and elliptic-curve cryptography (ECC) [6] have been developed, based on some well-known mathematical difficulties, including factorization of a large number in RSA and a discrete logarithm in DH. Those public key exchanges have become the foundation of today's cryptographic standards to establish a shared key for data encryption and decryption with symmetric cryptographic algorithms such as triple data encryption algorithm or TDEA [7] and advanced encryption standard or AES [8]. Those standard public key algorithms and the symmetric cryptographic algorithms form the standards of the transport layer security or TLS.

The security of these standards is at risk due to recent advancements in quantum computing. For instance, Shor's algorithm (1994) [9], provides an exponential speed up in breaking current public key exchange standards but was largely theoretical as there was no quantum computing hardware available. Google in September 2019 declared quantum supremacy based on its 54-qubit quantum computer called "Sycamore" — an achievement of a milestone in quantum computing development [10]. On the other hand, Grover (1996) invented a new quantum search algorithm called Grover's algorithm [11], which solves the unstructured search problem of size n in $O(\sqrt{n})$ queries, while any classical algorithm needs O(n) queries. This speedup requires the key length of standard AES to be raised from 128 bits up to 256 bits with true randomness. Therefore, the commercial availability of quantum computers poses a major existential threat to today's cryptographic standards, especially to public key exchanges, which shakes the foundations of contemporary information security.

Experts from academia and industry are heavily investing into researching potential candidates for public key exchange algorithms that can withstand attacks from quantum computing. From a mathematical perspective, candidates of post-quantum cryptographic (PQC) algorithms (see NIST website for Post-Quantum Cryptography) have been selected as PQC candidates by NIST. These algorithms can be classified into two categories: digital signature and key encapsulation mechanism (KEM). Within PQC KEM, there are Lattice-based such as NTRU [12] and Ring-Learning With Error (R-LWE) [13], code-based such as McEliece encryption system [14, 15] and random linear code encryption scheme (RLCE) [16], Multivariate [17] and supersingular isogeny Diffie-Hellman (SIDH) [18]. The security of different PQC algorithms is based on different computational difficulties, especially the hardest NP-problem such as the shortest vector problem (SVP) in lattice-based, error correcting problem in code-based, etc. NIST currently is in round 3 review of those candidates.

At the same time, quantum key distribution (QKD) has been proposed as an alternative solution for key distribution, using physical quantum systems, e.g., photons, as information carriers to achieve the theoretical security from the laws of quantum physics. QKD has been widely investigated since the first proposal by Bennett and Brassard in 1984 [19]. More recently, Xu et al (2019) [20] made a more complete review of its security analysis over the protocols, implementations, signal sources, and detections. Although QKD is proven to be semantically secure, its physical and technical limitations, as well as its costs and incompatibility to the existing infrastructure, render the technology impractical for commercial mass applications.

The work proposed in this paper may offer an avenue to a third alternative to PQC and QKD approaches for solving the key exchange problem. We propose to extend classical Shannon perfect secrecy from OTP's bijections with XOR operations over $GF(2^n)$ to the entire symmetric group $S_{2^n}$ over a field $GF(2^n)$ and explore its representations in a $DHS(2^n)$ defined over a computational basis $\{|0\rangle, |1\rangle, \ldots, |2^n-1\rangle\}$. The generalized Shannon perfect secrecy not only retains the perfect secrecy property, but gains reusability. We conjecture that such an extension would allow us to securely distribute true random keys, for example those generated by quantum random number generators, over existing network infrastructure, while providing theoretical provable properties of perfect secrecy.

The rest of this paper is organized as follows: In Section 2, we introduce Hilbert space over a computational basis and more specifically focus on discrete Hilbert sphere (DHS); in section 3, we investigate the representation of classical XOR operations in DHS; in section 4, we extend Shannon perfect secrecy to DHS and finally we draw our conclusion at the end.

## II. DISCRET HILBERT SPACE

For an n-bit information system, its information space is defined by a finite field or Galois field $GF(2^n)$, i.e., a set of integer numbers from 0 to $2^n-1$. Over a $GF(2^n)$, Boolean algebra is the foundation for building today's computing systems, as well as all applications and communication mechanisms. However, quantum computers are built with linear algebra in Hilbert space. From classical computing to quantum computing, the computing algebra makes a paradigm shift from Boolean algebra over $GF(2^n)$ to linear algebra in Hilbert space over a complex field $\mathbb{C}$. In the work presented in this paper we investigate the central question: "can we possibly make a similar shift for the Shannon perfect secrecy from a field $GF(2^n)$ to a Hilbert space?".

For an n-qubit system in quantum computing, a Hilbert space is represented by a computational basis $\{|0\rangle, |1\rangle, \ldots, |2^n-1\rangle\}$, in a Dirac bra-ket notation, kets are referred to column unit vectors

$$|0\rangle = \begin{bmatrix} 1 \\ 0 \\ . \\ . \\ . \\ 0 \end{bmatrix}, |1\rangle = \begin{bmatrix} 0 \\ 1 \\ . \\ . \\ . \\ 0 \end{bmatrix}, \ldots, |2^n-1\rangle = \begin{bmatrix} 0 \\ 0 \\ . \\ . \\ . \\ 1 \end{bmatrix}. \quad (1a)$$

and bras are referred to row unit vectors

$$\langle 0| = [\ 1\ 0\ \ldots\ 0], \ldots, \langle 2^n-1| = [\ 0\ 0\ \ldots\ 1] \quad (1b)$$

This basis in Eq. (1a) or Eq. (1b) is also known as a canonical basis. When using a lower-case letter within a bra or ket here, we are referring to the computational basis. The computational basis is orthonormal, so

$$\langle i | j \rangle = \delta_{ij}, \text{ for i, j} = 0, 1, 2, \ldots, 2^n-1 \quad (2)$$

where the Kronecker delta $\delta_{ij}$ in Eq. (2) takes 0 when $i \neq j$ and 1 when $i = j$. A system state of n-qubits generally is in a superposition state

$$|\psi\rangle = \sum_j \alpha_j |j\rangle \quad (3)$$

and the probability amplitudes, i.e., coefficients in Eq. (3), are required to meet

$$\sum_j |\alpha_j|^2 = 1 \quad (4)$$

where summation in Eqs. (3) and (4) runs from 0 to $2^n-1$ and the probability amplitudes are defined over a complex field $\mathbb{C}$: $\alpha_j \in \mathbb{C}$. Eq. (4) demonstrates that a superposition state is just a point on an abstract unit sphere called Hilbert sphere and is represented by a set of complex coefficients $\alpha_j$ with j=0, 1, 2, ..., $2^n-1$. Quantum computing often visualizes Qubit superposition through a special Hilbert sphere called a "Bloch sphere for single qubits" [21].

In quantum computing, operations on qubits can be described by quantum logic gates, such as the Hadamard gate. Mathematically, quantum logic gates are represented by $2^n \times 2^n$ unitary reversible matrices due to the constraint from Eq. (4). Unlike Boolean logic gates, quantum logic gates require $2^n$ inputs and $2^n$ outputs and must be reversible, hence their description by $2^n \times 2^n$ square matrices.

It is worth to emphasize that a $2^n \times 2^n$ square unitary matrix is a logical representation of an n-qubit quantum gate in its computational basis. The linear algebra defined in that computational basis can be logically used to express the corresponding quantum gate operation on its state of the system. That means, a gate operation can be logically implemented within the mathematical framework of Complex linear algebra. The logical implementation would also correctly show the behavior of the quantum system, but it would not demonstrate the parallel-computation advantages arising from the state superpositions of the physical system described by the same mathematics. These observations can be summarized as follows: for quantum computing, we need to have physical implementations of quantum gates to gain the parallel-computation advantages; however for quantum communications, logical implementations of quantum gates would allow us to take the advantages of the laws of quantum mechanics for communication security, logically not physically. This logic does not contradict the basic characteristics of a real physical quantum system where the system has discret states of quantized energy levels but correctly reflects the fundamental attributes of information systems that have discrete states of system information. Therefore, we may well consider and describe classical information systems (based on bits) as logical quantum systems, allowing us to extend the Boolean algebra over finite field to linear algebra over a computational basis.

The idea proposed in this paper is inspired from the work by Buniy, Hsu and Zee (2005) who proposed a concept to discretize the Hilbert sphere for qubits [22]. In particular, we recognized that there exists a particular discrete Hilbert sphere over a Galois field GF(2) that restricts probability amplitudes $\alpha_j \in [0, 1]$ and for which the discrete Hilbert sphere consists of $2^n$ states from its computational basis. In this superposition space described by the DHS, each state allows to be swapped to another state, but not in between states, under the constraint of Eq. (4). That means, DHS($2^n$) precisely reflects a classical n-bit system over the computational basis. Quantum mechanically, this superposition is represented by a permutation operator defined over the DHS($2^n$). There are total $2^n!$ unique permutation operators over the DHS with a computational basis consisting of $2^n$ unit vectors.

In group theory, the symmetric group $S_N$ is defined over a finite set of N symbols, whose elements are bijections from the set to itself. For a field GF($2^n$), the corresponding symmetric group is $S_{2^n}$ with $N = 2^n$. The order of this symmetric group is $2^n!$ and its all elements are permutations of the set. It is easy to notice that $2^n!$ permutation operators over DHS($2^n$) reflect $2^n!$ permutation elements from the symmetric group $S_{2^n}$ which tells that the DHS($2^n$) is the representation of the symmetric group $S_{2^n}$ in a Hilbert space. Fig. 1 symbolically illustrates their relationships. Inside the figure, 1a denotes the finite field GF($2^n$) of an n-bit system with $2^n$ decimal numbers as the states of the system; 1b expresses those states with $2^n$ circled dots on a unit Hilbert sphere over a computational basis $\{|0\rangle, |1\rangle, \ldots, |2^n-1\rangle\}$ and bijections of group actions are represented by permutation matrices within the DHS($2^n$); 1c demonstrates a physical n-qubit quantum computing system over a continuous Hilbert space or CHS($2^n$) where coefficients in Eq. (3) are defined over a complex field $\mathbb{C}$ under the condition in Eq. (4). From this representation, we can see a path from classical to quantum: GF($2^n$) $\rightarrow$ $S_{2^n}$ =DHS($2^n$) $\rightarrow$ CHS($2^n$), a continuous Hilbert space over a complex field $\mathbb{C}$. Within this context, the DHS($2^n$) for a classical n-bit system can be considered as a logical representation of CHS($2^n$) of a physical n-qubit system, clapsed at quantum measurements, and can also serve as a mathematical bridge for an extension of classical Shannon perfect secrecy from GF($2^n$) to Hilbert space DHS($2^n$). In the next section, we will first explore the expression of OTP XOR operations in this DHS($2^n$), and thereafter extend it to the entire symmetric group with this discrete Hilbert space.

### III. Shannon Perfect Secrecy in a Discrete Hilbert Space: OTP

For a classical n-bit system, Shannon perfect secrecy for OTP relates to bijections from a field GF($2^n$) to itself with XOR operations. In this section we describe how we can represent such bijections over a DHS($2^n$) with a computational basis $\{|0\rangle, |1\rangle, \ldots, |2^n-1\rangle\}$.

Consider a plaintext m and a random key k over GF($2^n$). The ciphertext c can be simply written as

$$c = k \oplus m \quad (5)$$

In a Hilbert space, an XOR operator, i.e., k$\oplus$, can be written as $\hat{X}_k$ with a subscript k $\in [0, 2^n-1]$. There is a total of $2^n$ unique XOR operators in a DHS($2^n$). We can express above Eq. (5) quantum mechanically as:

$$\hat{X}_k |m\rangle = |(k \oplus m)\rangle = |c\rangle, \quad (6)$$

that is, $\hat{X}_k$ operation in Eq. (6) transforms a state $|m\rangle$ into another state $|c=(k\oplus m)\rangle$ which is another basis vector in DHS($2^n$). Based on that, we can calculate the matrix element of $\hat{X}_k$ in a computational basis $\{|0\rangle, |1\rangle, \ldots, |2^n-1\rangle\}$ as follows

$$(\hat{X}_k)_{ij} = \langle i| \hat{X}_k |j\rangle = \langle i | j'=(k \oplus j)\rangle$$
$$= \delta_{i(j'=k\oplus j)}, \quad (7)$$

where subscripts i and j are from 0 to $2^n$-1. The operator $\hat{X}_k$ in Eq. (7) is mathematically described by a $2^n \times 2^n$ binary permutation matrix. The Kronecker delta indicates that the XOR operation with a random key k performs column permutations of the identity matrix by XORing the key k with column index j to give its new column index j′. The permutation matrix is created by performing this column permutation for all rows. It is also clear from Eq. (7) that all XOR permutation matrices are symmetric and binary so their inverse and transpose are the same as themselves. For the example where n = 3 bits, the computational basis consists of 8 basis vectors $\{|0\rangle, |1\rangle, |2\rangle, |3\rangle, |4\rangle, |5\rangle, |6\rangle, |7\rangle\}$, a bijection encryption with the XOR operator $\hat{X}_3$ can be directly written into a matrix multiplication within the computational basis

$$\hat{X}_3 |m\rangle = |c\rangle \rightarrow \begin{bmatrix} 0 & 0 & 0 & 1 & 0 & 0 & 0 & 0 \\ 0 & 0 & 1 & 0 & 0 & 0 & 0 & 0 \\ 0 & 1 & 0 & 0 & 0 & 0 & 0 & 0 \\ 1 & 0 & 0 & 0 & 0 & 0 & 0 & 0 \\ 0 & 0 & 0 & 0 & 0 & 0 & 0 & 1 \\ 0 & 0 & 0 & 0 & 0 & 0 & 1 & 0 \\ 0 & 0 & 0 & 0 & 0 & 1 & 0 & 0 \\ 0 & 0 & 0 & 0 & 1 & 0 & 0 & 0 \end{bmatrix} \begin{bmatrix} |0\rangle \\ |1\rangle \\ |2\rangle \\ |3\rangle \\ |4\rangle \\ |5\rangle \\ |6\rangle \\ |7\rangle \end{bmatrix} = \begin{bmatrix} |3\rangle \\ |2\rangle \\ |1\rangle \\ |0\rangle \\ |7\rangle \\ |6\rangle \\ |5\rangle \\ |4\rangle \end{bmatrix} \quad (8)$$

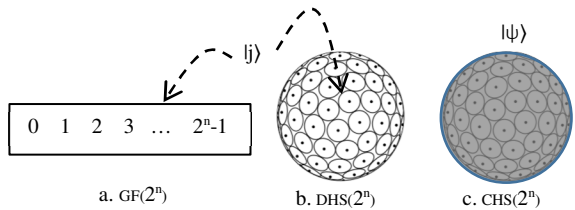

Figure 1. The plot illustrates a path from a finite field GF($2^n$) to a continuous Hilbert Space CHS($2^n$). Inside this figure, a finite field of an n-bit system is shown on the left, corresponding discrete Hilbert and continuous Hilbert spheres are shown in the middle and on the right. The transparent Hilbert sphere in DHS($2^n$) tells that superposition states can only be on those circled dots not in between two dots. The semi-transparent unit sphere in CHS($2^n$) indicates continuous superposition states defined over a complex field in an n-qubit quantum computing system.

Eq. (8) demonstrates that the computational basis is transformed to a cipher basis by the permutation matrix. A plaintext "2" in Eq. (8) is denoted by a state $|2\rangle$ then encrypted with $\hat{X}_3$ into a cipher state $|1\rangle$, referred to a ciphertext "1".

The reverse transformation for decryption performs in the same way due to the symmetric characteristics

$$\hat{X}_3 |c\rangle = |m\rangle \rightarrow \begin{bmatrix} 0 & 0 & 0 & 1 & 0 & 0 & 0 & 0 \\ 0 & 0 & 1 & 0 & 0 & 0 & 0 & 0 \\ 0 & 1 & 0 & 0 & 0 & 0 & 0 & 0 \\ 1 & 0 & 0 & 0 & 0 & 0 & 0 & 0 \\ 0 & 0 & 0 & 0 & 0 & 0 & 0 & 1 \\ 0 & 0 & 0 & 0 & 0 & 0 & 1 & 0 \\ 0 & 0 & 0 & 0 & 0 & 1 & 0 & 0 \\ 0 & 0 & 0 & 0 & 1 & 0 & 0 & 0 \end{bmatrix} \begin{bmatrix} |3\rangle \\ |2\rangle \\ |1\rangle \\ |0\rangle \\ |7\rangle \\ |6\rangle \\ |5\rangle \\ |4\rangle \end{bmatrix} = \begin{bmatrix} |0\rangle \\ |1\rangle \\ |2\rangle \\ |3\rangle \\ |4\rangle \\ |5\rangle \\ |6\rangle \\ |7\rangle \end{bmatrix} \quad (9)$$

Eq. (9) demonstrates that the cipher basis is transformed back to the computational basis by the same permutation matrix. The cipher state $|1\rangle$ in Eq. (9) is decrypted back to a state $|2\rangle$ referred to plaintext "2". There exists a total of $2^3 = 8$ XOR bijections among the entire permutations $2^3! = 40320$ from the computational basis of a 3-bits system to itself. It should be noticed that Eqs. (8) and (9) are only examples to demonstrate how OTP encryption and decryption work in its corresponding computational basis in a scope of quantum computing.

A generic OTP pad may contain a very long stream of bits such as n = 1,000,000 bits. It is impractical to create a permutation matrix with $2^n \times 2^n$ in this case. A common practice is to segment the longer bit-string into words with a relative small fixed length r per word such as r = 8-bits or r = 16-bits. For an 8-bit word, the total XOR permutation matrices are $2^8 = 256$ and $\hat{X}_i$ with i from 0 to 255. After this segmentation, the huge $2^n \times 2^n$ permutation matrix is simplied to a block diagonal matrix and can be expressed with a set of r-bit permutation matrices $\{\hat{X}_i\}$ or a pad of XOR permutation matrices.

The encryption, as well as decryption, is expressed as a permutation matrix multiplying with the computational basis. That means, a bijection in a OTP transformation is generally represented by a corresponding symmetric permutation matrix multiplication with a computational basis. We can conclude that total $2^n$ XOR symmetric permutation matrices represent the entire bijections of a classical Shannon perfect secrecy.

We note that the limitation of one-time use can be also expressed and interpreted quantum mechanically. For instance, given two XOR operators $\hat{X}$ and $\hat{X}'$ and performing

$$\hat{X}\hat{X}' | m \rangle = | x \oplus (x' \oplus m) \rangle = | x' \oplus (x \oplus m) \rangle = \hat{X}'\hat{X} | m \rangle$$

then the above equation can be rewritten in the form of a commutator:

$$[\hat{X}, \hat{X}']| m \rangle = (\hat{X}\hat{X}' - \hat{X}'\hat{X})| m \rangle \equiv 0 \quad (10)$$

Therefore, XOR operators are commutable in a DHS($2^n$) defined by the computational basis. From quantum mechanics, Eq. (10) shows that XOR operators intrinsically share the same set of eigenstates although each makes different transformation. It is this commutability or shared eigenstates which restricts OTP to no more than one-time use to maintain the perfect secrecy.

IV. SHANNON PERFECT SECRECY IN A DISCRETE HILBERT SPACE: QUANTUM PERMUTATION PAD

More generically, Shannon perfect secrecy states a set of bijections from plaintext space to ciphertext space with a truly random key from a key space. OTP is a specific case of Shannon perfect secrecy with XOR as the bijection operation. Inside of the symmetric group $S_{2^n}$, there are total $2^n! = 2^n (2^n-1)!$ bijections, which means that $(2^n-1)!$ sets of $2^n$ bijections (1-to-1 mapping) exist between plaintext space and ciphertext space. We can also say that the 1-to-1 mapping from OTP is degenerate and the degeneracy is $(2^n-1)!$. The extension of the Shannon perfect secrecy to the entire symmetric group $S_{2^n}$ or in a DHS($2^n$) is a decoupling process of degeneracy from 1-to-1 mapping to $1-(2^n-1)!-1$ mappings, that means, existing exact $(2^n-1)!$ pathes between any pair of plaintext and ciphertext.

A symmetric group $S_{2^n}$ over a set of items $\{0, 1, 2, \ldots, 2^n-1\}$ can be accurately described by a DHS($2^n$) defined over a computational basis $\{|0\rangle, |1\rangle, \ldots, |2^n-1\rangle\}$, within this DHS($2^n$), all permutation operators are represented by $2^n \times 2^n$ binary permutation matrices.

Let $\hat{P}_k$ denote a permutation operator representing a bijection from the symmetric group with k=1, 2, …, $2^n!$. Only a very small portion of bijections can be specifically expressed by some mathematic operators, such as XOR or addition operators, while the vast majority of permutation operators cannot be specifically expressed with mathematical operators. Therefore, we will generally use permutation operator $\hat{P}$ to

indicate a bijection operation. The four major features of symmetric group and permutation operators are:

- The symmetric group $S_{2^n}$ consists of all $2^n!$ bijection mappings from the computational basis $\{|0\rangle, |1\rangle, \ldots, |2^n-1\rangle\}$ to itself with equally likely probability.

- A bijection is represented by a permutation operator $\hat{P}_k$, a $2^n \times 2^n$ unitary binary permutation matrix in a DHS($2^n$) over a computational basis $\{|0\rangle, |1\rangle, \ldots, |2^n-1\rangle\}$, and k =1, 2, …, $2^n!$. A unitary permutation matrix is a logical representation of a quantum permutation gate.

- Each permutation matrix P has a reverse to be its transpose so $P^{-1} = P^T$ so $PP^{-1} = PP^T = 1$. This feature has significant practical value as it allows for fast and energy efficienct transformations of the data,

- In general, two permutation matrices are non-commutable: $[P, P'] \neq 0$ in a DHS($2^n$) with n > 1.

We would like to highlight two of these features for further discussion: the first is the key space from a finite field GF($2^n$) with $2^n$ states to the entire symmetric group $S_{2^n}$ with $2^n!$ elements, which maximizes the intactability for security. Even for a small n = 8 bits, $2^8! = 10^{507}$ is possibly beyond any computing capability to brute force. At the same time, the probability distribution extends the Shannon perfect secrecy for transformations over the entire symmetric group.

The second noteworthy feature is the non-commutability, or generalized uncertainty principle with n > 1. When n = 1, the DHS($2^1$) is identical to the GF(2) where two permutation matrices are commutable. Although there exists the non-commutativity with n = 2, where DHS($2^2$) is associated with a symmetric group $S_4$, a solvable group or factorizable group based on Galois theory, it is recommended that n should be larger than 2 to make the group unslovable. Under this consideration, the non-commutativity can be quantum mechanically interpreted that permutation operators generally do not have a shared eigenspace over a DHS($2^n$). That means, transformations by permutation matrix P on its computational basis show general nonlinear behavior over the GF($2^n$). P operates on a plaintext state $|m\rangle$ to produce a cipher state $|c\rangle$,

$$P|m\rangle = |c\rangle,$$

then the Eve intercepts the cipher state $|c\rangle$ and plays a test permutation operation with P'

$$P'|c\rangle = P'P|m\rangle = P''|m\rangle = |c'\rangle$$

which shows that a multiplication of two permutation matrices P and P' produces a new permutation matrix P″ so a test attack equals a new permutation transformation and turns the cipher state $|c\rangle$ into a new cipher state $|c'\rangle$. Only in one case out of the $2^n!$ possible choices would yield the desired outcome, that is, $P'' = P'P = 1$ which reveals the original plaintext state $|m\rangle$, where P' is the transpose of the randomly selected P. However, under the assumption that commutivibility exists within the group of bijections such as XORs in OTP, then a permutation transformation shows a linear behaviour over GF($2^n$) and the effect of the transformation could be eliminated once the transformation is repeatedly used even without knowing the actual key. Therefore, this non-commutativity over the entire symmetric group generally makes the permutation transformations to be nonlinear, which allows a QPP to be reusable without invalidate the perfect property.

Based on the above discussions, the extended Shannon perfect secrecy can be expressed as: **an intercepted ciphertext provides no information about the plaintext, or the pad QPP**.

Let's take the same example of 3-bit system with a computational basis $\{|0\rangle, |1\rangle, |2\rangle, |3\rangle, |4\rangle, |5\rangle, |6\rangle, |7\rangle\}$ to demonstrate how a generic permutation encryption works. Instead of using an XOR operator, we randomly choose a permutation matrix $\hat{P}_k$ from $2^3!=40320$ matrices, and perform a matrix ($\hat{P}_k$) multiplication with the computational basis:

$$\hat{P}_k |m\rangle = |c\rangle \rightarrow \begin{bmatrix} 0 & 1 & 0 & 0 & 0 & 0 & 0 & 0 \\ 0 & 0 & 0 & 0 & 1 & 0 & 0 & 0 \\ 0 & 0 & 1 & 0 & 0 & 0 & 0 & 0 \\ 0 & 0 & 0 & 0 & 0 & 1 & 0 & 0 \\ 0 & 0 & 0 & 1 & 0 & 0 & 0 & 0 \\ 1 & 0 & 0 & 0 & 0 & 0 & 0 & 0 \\ 0 & 0 & 0 & 0 & 0 & 0 & 0 & 1 \\ 0 & 0 & 0 & 0 & 0 & 0 & 1 & 0 \end{bmatrix} \begin{bmatrix} |0\rangle \\ |1\rangle \\ |2\rangle \\ |3\rangle \\ |4\rangle \\ |5\rangle \\ |6\rangle \\ |7\rangle \end{bmatrix} = \begin{bmatrix} |1\rangle \\ |4\rangle \\ |2\rangle \\ |5\rangle \\ |3\rangle \\ |0\rangle \\ |7\rangle \\ |6\rangle \end{bmatrix} \quad (11)$$

Eq. (11) shows that the computational basis is transformed to a cipher basis with the selected permutation matrix. A plaintext "3" in Eq. (11) is encrypted into a ciphertext "5". And the decryption is to perform $\hat{P}_k$'s transposed permutation matrix multiplication with the cipher basis

$$\hat{P}_k^\dagger |c\rangle = |m\rangle \rightarrow \begin{bmatrix} 0 & 0 & 0 & 0 & 0 & 1 & 0 & 0 \\ 1 & 0 & 0 & 0 & 0 & 0 & 0 & 0 \\ 0 & 0 & 1 & 0 & 0 & 0 & 0 & 0 \\ 0 & 0 & 0 & 0 & 1 & 0 & 0 & 0 \\ 0 & 1 & 0 & 0 & 0 & 0 & 0 & 0 \\ 0 & 0 & 0 & 1 & 0 & 0 & 0 & 0 \\ 0 & 0 & 0 & 0 & 0 & 0 & 0 & 1 \\ 0 & 0 & 0 & 0 & 0 & 0 & 1 & 0 \end{bmatrix} \begin{bmatrix} |1\rangle \\ |4\rangle \\ |2\rangle \\ |5\rangle \\ |3\rangle \\ |0\rangle \\ |7\rangle \\ |6\rangle \end{bmatrix} = \begin{bmatrix} |0\rangle \\ |1\rangle \\ |2\rangle \\ |3\rangle \\ |4\rangle \\ |5\rangle \\ |6\rangle \\ |7\rangle \end{bmatrix} \quad (12)$$

in Eq. (12), the transposed permutation matrix transforms the cipher basis back to the computational basis. The ciphertext "5" in Eq. (12) is decrypted back to the plaintext "3". Total $(2^3-1)! = 5040$ permutation matrices can equally likely encrypt "3" into "5" and decrypt "5" back to "3". That means, the 1-to-1 mapping in OTP becomes 1-5040-1 mappings here. For a QPP with M permutation matrices $\{P_i(2^n \times 2^n)$, n is the number of bits per word and i=1, 2, .., M$\}$, the encoding and decoding processes are the same for each n-bit word and are simply repeated M times for a block of M words.

There are different ways to randomly select permutation matrices with a given random bit string. The key scheduling algorithm in RC4 can be used to shuffle states of GF($2^8$) for 8-bit computational basis. It should be not difficult to modify it for any size. Another shuffling algorithm is the Fisher-Yates algorithm. The following is an example of pseudo codes to

demonstrate how a QPP can be selected with a random string of bits with the Fisher Yates shuffle algorithm:

```
-- state array S[2ⁿ] → a permutation matrix P[2ⁿ][ 2ⁿ]
-- initialize P[2ⁿ][ 2ⁿ] to all zeros
for i from 0 to 2ⁿ -1
    S[i] = i
-- input random key k[N] with N =2ⁿ
for i from 2ⁿ –1 downto 1 do
   j = k[i]
   swap S[j] and S[i]
for i from 0 to 2ⁿ -1
    P[i][S[i]] = 1
```

the above psudeo-code shows how a single permutation matrix is randomly selected with an input random key of $n2^n$ bits to be pre-shared on both sender and receiver. For a QPP with M permutation matrices, the total key length can be as long as $M(n2^n)$ bits. However, the total length can be shorter than that based on a practical security requirement. For a typical logical implementation, n would be 8 and M would be 16. Then it needs a random stream of bits as long as 32,768 bits to create a QPP with 16 permutation matrices.

A QPP can thus be considered as a OTP extension with resuability for information transformation over a $DHS(2^n)$. In contrast to OTP where encryption is carried out word-by-word by XORing with the pad, a QPP does permutation matrix multiplication word-by-word, simply replacing XOR with permutation matrix multiplications. This simple operation replacement does a fundamental shift of the Shannon perfect secrecy from $GF(2^n)$ to $DHS(2^n)$ and enable the pad to be reusable without invalidating the feature of perfect secrecy, directly benefiting from the computing algebra shifting from Boolean algebra to linear algebra. Another major difference between OTP and QPP is the Shannon information entropy. An OTP with M words of n-bits has $M*n$ bits of entropy but a QPP has $M*\log_2(2^n!) = M*n + M*\log_2(2^n-1)!$ bits of entropy, much much larger than $M*n$. Therefore, we receive three significant benefits from the extension of Shannon perfect secrecy to $DHS(2^n)$: perfect, reuable and huge entropy.

If we perform a thought experiment of quantum key distributions with an n-qubit physical system, not single qubit as in a traditional QKD, we may have to use a physical n-qubit permutation gate or corresponding Hadamard gate on both sender and receiver and initialize them with a pre-shared secret before starting key distributions because it is impossible to randomly select a permutation gate from $2^n!$ permutation gates to encode at the sender and to measure at the receiver. One could image the practical difficulty to realize this thought experiment. However, the logical implementation of this thought experiment is straightforward as what this paper proposes. One may ask what the reason is for the pre-shared secret, the answer is to ensure the trust like today's QKD does. Therefore, QPP could be the ideal alternative to traditional QKD for point-to-point true random key distributions over the existing internet without constrainsts of a physical QKD.

In classical cryptographies, a substitution box or S-box is widely used for a nonlinear transformation such as in DES and AES ciphers. Heys and Tavers [23] have shown that using randomly selected large S-boxes and suitable linear transformations can largely improve the security of product ciphers and offer good resistant to the differential and linear cryptanalysis. S-boxes are essentially bijections from a $GF(2^n)$ to itself, further converted to permutation matrices, i.e. a QPP. It is our interests to further investigate a random selected QPP to be combined with algorithms with diffusion capabilities such as AES and offer quantum safe cryptography for encryption of biased plaintexts, especially a light-weight cryptography for internet of things (IoT).

V. CONCLUSION

In this paper we propose an extension of classical Shannon perfect secrecy for One-Time Pad to Discrete Hilbert space $DHS(2^n)$ over the entire symmetric group $S_{2n}$. We show that QPP operations within this Hilbert space are represented by the complete set of permutation matrices over an n-bit computational basis. In this context, the bijections of the classical 1-to-1 mappings between plaintext, pad and ciphertext are equally likely decomposed into $1-(2^n-1)!-1$ mappings, so the perfect secrecy is maintained by this extension. In addition, non-commutability within permutation group is generally expressed by the generalized uncertainty principle that allows repeated uses of the QPP pad within this extension, making the Shannon perfect secrecy not only for textbooks but also for practical industry applications. One desired application for our extension is to distribute true random keys from one place to any other place without limitations as what today's QKD does, which immediately allows us to be able to defend the incoming quantum computing threats. Future work will first develop this extended Shannon perfect Secrecy as alternative of QKD for true random number distributions.

ACKNOWLEDGMENTS

We want to acknowledge Ken Dobell for his editorial assistance.